\documentclass[a4paper,12pt]{article}
\usepackage[utf8]{inputenc}
\usepackage{cancel}
\usepackage{ulem}
\usepackage{amsfonts}
\usepackage{amssymb}
\usepackage{graphicx}
\usepackage{amsmath}
\usepackage{enumerate}
\usepackage{mathtools}
\usepackage{subfig}
\usepackage{color}
\usepackage{tikz}
\usepackage{float}
\usepackage{here}
\usepackage{cite}
\usepackage{mathrsfs}
\usepackage{float,epsfig}
\usepackage{dcolumn}
\usepackage{graphicx}
\usepackage{bm}
\usepackage{amsmath,amssymb,amsthm}
\usepackage[colorlinks=true,linkcolor=blue,citecolor=red]{hyperref}
\usepackage{multirow}

\usetikzlibrary{calc}
\newcommand{\be}{\begin{equation}}
\newcommand{\ee}{\end{equation}}
\newcommand{\bea}{\setlength\arraycolsep{2pt} \begin{eqnarray}}
\newcommand{\eea}{\end{eqnarray}}

\def\0{{\sst{(0)}}}
\def\1{{\sst{(1)}}}
\def\2{{\sst{(2)}}}
\def\3{{\sst{(3)}}}
\def\4{{\sst{(4)}}}
\def\5{{\sst{(5)}}}
\def\6{{\sst{(6)}}}
\def\7{{\sst{(7)}}}
\def\8{{\sst{(8)}}}
\def\sst#1{{\scriptscriptstyle #1}}

\makeatletter \@addtoreset{equation}{section}

\definecolor{lime}{HTML}{A6CE39}

\begin{document}

\title{{\normalsize \textbf{On   Inflation and  Axionic  Dark Matter in a Scaled
Gravity }}}
\author{ {\small  A. Belhaj$^{1}$\footnote{a-belhaj@um5r.ac.ma}, S. E. Ennadifi$^{2}$\footnote{Ennadifis@gmail.com},  M. Lamaaoune$^{1}$\footnote{lamaaoune2944massi@gmail.com}\thanks{
{\bf Authors in alphabetical order}.} \hspace*{-8pt}} \\
{\small $^1$ D\'{e}partement de Physique, Equipe des Sciences de la
mati\`ere et du rayonnement, ESMaR}\\
{\small Facult\'e des Sciences, Universit\'e Mohammed V de Rabat, Rabat,
Morocco} \\
{\small $^2$ LPHE-MS, Facult\'e des Sciences, Universit\'e Mohammed V de
Rabat, Rabat, Morocco} }
\maketitle

\begin{abstract}
Motivated by the modified gravity theories $F(R)\neq R$ and inflationary  physics, we  first propose and investigate  an inflation model in a scaled
gravity $F(R)=R\,+\beta R$, where $\beta $ is    a dimensionless scaling
parameter. The latter is also implemented in  a particular  potential  $V(\phi )=M^{4}\left[ 1-\cos \left( \frac{\phi }{\mu}\right)
^{\beta }\right] $  being considered to drive the inflation   via a parameter coupling scenario.  Using the slow-roll
approximations, the gravity scale parameter $\beta $ is approached  with respect to the range of the   associated  computed cosmological
observables $n_{s}$ and $r$ according to the recent Planck and BICEP/Keck
data. Then, we   discuss  the  axionic dark matter in the suggested gravity  model by considering 
the case where the inflaton  is taken to be identified with  an  axion-like
field $\phi =f_{a}\theta $ with  the decay constant $f_{a}=\mu$.  Referring to the known data, the underlying inflation scale $M$ is
constrained to be much lower than the corresponding  axion scale  $M\ll f_{a}$.
\end{abstract}

%


\newpage

\section{Introduction}

Recently, inflationary models have been studied in depth by considering
many theories of gravity \cite{r0,r00,r1,r2,r3,r4,r5,r6,r7,r8,r9,r11,r12,r13,r14,r15,M1,M2,r16,r17,r18,r19,r23,r24,r25,r26,r27,M3}. Certain models have opened
new ways to understand the evolutionary phenomena of the Universe, such as
the horizon, the flatness and the problems of large structures \cite{r6, r7,
r8}. Various gravity theories have been suggested by exploiting  single and
multiple scalar fields through   potentials in order to determine  the standard
cosmological observables in the context of general relativity (GR). The
simplest models involve a single field that is considered as  the main
component of inflation. This scalar field has been approached using many
theories, including higher dimensional supergravity ones\cite{r13,r14,r15,M1,M2,r16,r17,r18}.  Concretely, advances in string theory and related topics have 
been exploited to develop inflationary models based on D-brane physics using
the Randall-Sundrum II (RS-2) mechanism. 
In this way, interesting scalar
potentials in the presence of the stringy parameters such as the brane
tension have been analyzed where the stringy corrections of the involved
quantities have been obtained \cite{r13,r14,r15,M1,M2}. An examination
reveals that the compactification of the superstring models and M-theory
could generate many scalar fields being derived from the geometric
deformations of the metric and the non-trivial tensor fields defining the
stringy moduli space. These scalar fields have been explored to confront  predictions with observations from the cosmological microwave
background (CMB) and the Planck experiments\cite{r20,r21,r22}.\newline
More recently, modified models of GR have been explored  providing
interesting inflationary results. The most discussed ones are $F(R)$ modified
gravity theories where $R$ denotes the Ricci scalar. They are widely studied
by dealing with several scalar potentials\cite{r23,r24,r25}. Based on  the
slow-roll  analysis, the associated spectral index $n_s$ and the tensor/scalar
ratio $r$ have been determined with particular values of the number of
e-folds required by the observational results. These modified gravity models
have been extended by adding other quantities, including the trace $T$ of
the stress-energy tensor. The resulting $F(R,T)$ theory of the gravity is
intensively investigated allowing the realization of  inflationary
models\cite{T1,T2,T3,T4,T5,T6}. Some of them have been  motivated  by the study of dark
energy (DE) \cite{d1,d2}.  It has been pointed out that this substance can be explored to
provide explanations for the accelerated aspect of the expansion of the
Universe. Specifically, the modified gravity theories could generate
dynamical phenomena which can be assimilated to contributions associated with
DE. These behaviors exceed the cosmological constant in the context of GR.

It has been  remarked  that the form of the scalar potential can be of crucial
importance in the building of inflationary models arising from various
theories including  superstring models and M-theory. The choice of the form of
the potential generally depends on motivations supported by known models
like the standard model (SM) of particle physics \cite{sm1,sm2}. It has been noted
that famous examples are the chaotic inflation potential and the minimal
supersymmetric standard model (MSSM) inflation potential. In addition to
these known models, other types of scalar potentials have been discussed in
the context of dark matter (DM) \cite{dm1}.    In connections with inflation activities, several  DM  candidates  have been proposed and  considered.  However, it  has been suggested that axions  could be considered   as  relevant  DM candidates  via   certain vacuum 
fluctuations  during (or at the end)  of inflation  \cite{dm1,dm2,dm3,dm4}. It has been  shown that these scalar  field  have been introduced in different  ways.   One of them is associated with the CP problem via the Peccei and  Quinn symmetry in the quantum chromodynamics (QCD) context.  These axions are called QCD axions \cite{dm5,dm6}. Other types of axions appear naturally in string theory  dealing with higher dimensional objects like strings  and  branes in extra dimensional space-times. In this way,  the associated scalars are called axion-like particles derived from topological and geometrical contributions of  the internal compact geometries associated with extra dimensions\cite{dm1,dm7,dm8}. Alternatively, axions could appear also in Chern-Simons  (CS) interactions with  gravity  producing axions-CS gravity \cite{dm9,dm10}.

In the examination of inflation parameters, one should distinguish two
categories. The first one contains the parameters of the gravity sector.
However, the second one involves the matter  parameters including the dark sector contributions. A close
inspection reveals that one can follow two different  inflation scenarios where such
parameters are linked  or not.

Motivated by the modified gravity theories $F(R)\neq R$ and inflationary
physics, we  first propose and study   an inflation model in a scaled
gravity $F(R)=R\,+\beta R$, where $\beta $ is  a   dimensionless scaling
parameter. The latter is implemented in the special    potential  $V(\phi )=M^{4}\left[ 1-\cos \left( \frac{\phi }{\mu}\right)
^{\beta }\right] $ being  considered to drive the inflation   by means of a   parameter coupling scenario.  Using the slow-roll
approximations, the gravity scale parameter $\beta $ is approached  with respect to the range of the corresponding  computed  cosmological
observables $n_{s}$ and $r$ according to the recent Planck and BICEP/Keck
data.  After that, we    investigate   axionic DM in the suggested scaled   gravity  model by discussing 
the case where the inflaton  is identified with  an axion-like
field $\phi =f_{a}\theta $ with  the decay constant $f_{a}=\mu$.  Considering  known data, the underlying inflation scale $M$ is
constrained to be much lower than the associated axion scale  $M\ll f_{a}$.

The organisation of this paper is as follows. In section 2, we present
inflation calculations in a scaled gravity. In section 3, we investigate
parameter decoupling and coupling scenarios  for a scalar  potential
supported by DM activities. In section 4, we discuss the corresponding  axionic DM investigation.  The last section is devoted to concluding remarks.

\section{ Inflation in a scaled gravity}

In this section, we reconsider the study of $F(R)$ gravity with a kinetic
coupling term in order to investigate the inflation scenarios and related
topics including DM. Indeed, we start by taking the following action 
\begin{equation}
S=\int dx^{4}\sqrt{-g}\left( \frac{F(R)}{16\pi }+\mathcal{L}_{m}\right)
\end{equation}%
where $F(R)$ is an arbitrary function of the Ricci scalar $R$ and $g$
denotes the determinant of the metric $g_{\mu \nu }$ \cite{T1,T2,T3,T4,T5,T6}. $\mathcal{L}_{m}$ represents the matter sector   given by 
\begin{equation}
\mathcal{L}_{m}=-\frac{1}{2}(g^{\mu \nu }-\omega ^{2}G^{\mu \nu }) \bigtriangledown
_{\mu }\phi  \bigtriangledown _{\nu }\phi -V(\phi )
\end{equation}%
where $G_{\mu \nu }$ indicates the Einstein tensor and $\omega $ is a
positive parameter having the dimension of the inverse mass scale \cite{r28,r29,r30,r31,r32}.  The positive sign
of the kinetic coupling term $\omega ^{2}G^{\mu \nu }\partial _{\mu }\phi
\partial _{\nu }\phi $ is needed to remove the ghost in the studied model 
\cite{sm1,sm2}. This term has been implemented in the matter sector in
order to reduce the tensor-to-scalar ratio $r$ needed  to establish bridges with CMB
observations. In the matter sector, the relevant piece is the dynamical
scalar field $\phi $ controlled by a  potential $V(\phi )$.   Recently, many different $F(R)$  function forms  have been considered in inflation activities  in connection with many topics including  swampland criteria.  A particular emphasis has been put  on a rescaled Einstein-Hilbert  theory with  $F(R) \sim \alpha R$ where $\alpha$ is   a dimensionless  constant parameter  constrained by $0<\alpha <1$\cite{ref1,ref2,ref3}. Inspired by such activities, we would like to implement a $F(R)$  gravity parameter in order to establish a    coupling scenario  between matter and gravity  sectors, via the scalar potential.  In particular,  we  would like to identify a  parameter in the potential  with one of  $ F(R)$  function. The latter will be relevant in the inflation discussion.  Concretely,  we consider a scaled gravity described by the  following  function of $R$
\begin{equation}
F(R)=R\,+\beta R  \label{frt}
\end{equation}%
where $\beta $ is a dimensionless free parameter being independent of $R$.
This theory can be derived by considering the following scaling \begin{equation}
R\rightarrow (1+\beta )R  \label{R1}
\end{equation}%
producing a scaled gravity.   An other  possible justification for the use of such 
a gravity lies in the fact that it can be exploited to provide models  with linked parameters of matter and gravity  sectors   by means of the coupling scenario.
 In such a gravity, the previous action reduces 
to 
\begin{equation}
S=\int dx^{4}\sqrt{-g}\left( \frac{(1+\beta )R}{16\pi }-\frac{1}{2}\left(
g^{\mu \nu }-\omega ^{2}G^{\mu \nu }\right) \bigtriangledown _{\mu }\phi
\bigtriangledown _{\nu }\phi -V(\phi )\right).  \label{B1}
\end{equation}%
For the moment, it is worth noting that $\beta $ is different to $-1$ and $0$. Varying the
action given by  Eq.(\ref{B1}) with respect to the metric $g_{\mu \nu }$ and $\phi $, we
get the  equations of motion  which read as 
\begin{eqnarray}
(1+\beta )G_{\mu \nu }\,=\,8\pi (T_{\mu \nu }^{(\phi )} &+&\omega ^{2}\Theta
_{\mu \nu })  \label{Eq2} \\
\left( g^{\mu \nu }-\omega ^{2}G^{\mu \nu }\right) \bigtriangledown _{\mu
}\bigtriangledown ^{\mu }\phi &=&\frac{dV(\phi )}{d\phi}
\end{eqnarray}%
where $T_{\mu \nu }^{(\phi )}$ and $A_{\mu \nu }$ are given by 
\begin{eqnarray}
T_{\mu \nu }^{(\phi )} &=&\triangledown _{\mu }\phi \triangledown _{\nu
}\phi -\frac{1}{2}g_{\mu \nu }\triangledown _{\rho }\phi \triangledown
^{\rho }\phi +g_{\mu \nu }V(\phi ) \\
\Theta _{\mu \nu } &=&-\frac{1}{2}\triangledown _{\mu }\phi \triangledown
_{\nu }\phi R+2\triangledown _{\alpha }\phi \triangledown (_{\mu }\phi
R_{\nu }^{\alpha })+\triangledown ^{\alpha }\phi \triangledown ^{\beta }\phi
R_{\mu \alpha \nu \beta }  \notag \\
&+&\triangledown _{\mu }\triangledown ^{\alpha }\phi \triangledown _{\nu
}\triangledown _{\alpha }\phi -\triangledown _{\mu }\triangledown _{\nu
}\phi \square \phi -\frac{1}{2}(\triangledown \phi )^{2}G_{\mu \nu } \\
&+&g_{\mu \nu }[-\frac{1}{2}\triangledown ^{\alpha }\phi \triangledown
^{\beta }\phi \triangledown _{\alpha }\phi \triangledown _{\beta }\phi +%
\frac{1}{2}(\square \phi )^{2}-\triangledown _{\alpha }\phi \triangledown
_{\beta }\phi R^{\alpha \beta }].  \notag
\end{eqnarray}%
Taking $\beta =\omega =0$, for instance, we recover the usual standard field
equations. Using $g_{\mu \nu }$ and $\phi $ variations via the
Friedman-Lemaitre-Robertson-Walker (FLRW) metric, we obtain the equations of
motion 
\begin{eqnarray}
3(1+\beta )H^{2} &=&4\pi \dot{\phi}^{2}(1+9\omega ^{2} H^{2})+8\pi V(\phi ), \\
(1+\beta )(2\dot{H}+3H^{2}) &=&-4\pi \dot{\phi}^{2}(1-\omega ^{2} (2\dot{H}%
+3H^{2}+4H\ddot{\phi}\dot{\phi}^{-1}))+8\pi V(\phi ) \\
(\ddot{\phi}+3H\dot{\phi}) &+&3\kappa (H^{2}\ddot{\phi}+3H^{3}\dot{\phi}+2H%
\dot{H}\dot{\phi})=-V^{\prime }(\phi )
\end{eqnarray}%
where one has used the notations $^{\prime }=\frac{d}{d\phi }$ and $\,$ $%
\dot{}=\frac{d}{dt}$. It is recalled that $H$ denotes the Hubble parameter
defined by $H=\frac{\dot{ a}(t)}{a(t)}$ where $a(t)$ is the scalar factor.
To confront the proposed model with the observational data, one should
exploit the slow-roll analysis  by computing the associated parameters.
During the inflation phase, they are given by 
\begin{equation}
\epsilon =-\frac{\dot{H}}{H^{2}},\quad \eta =\frac{\dot{\epsilon}}{H\epsilon 
},\quad \kappa _{0}=12\pi \kappa \dot{\phi}^{2},\quad \kappa _{1}=\frac{\dot{%
\kappa _{0}}}{H\kappa _{0}}
\end{equation}%
being constrained by 
\begin{equation}
\epsilon ,\eta ,\kappa _{0},\kappa _{1}<<1.
\end{equation}%
In this way, the field equations of motion take the following simplified
forms 
\begin{align}
3H\dot{\phi}+9\kappa H^{3}\dot{\phi}=& -V^{\prime }(\phi ) \\
3(1+\beta )H^{2}=& 8\pi V(\phi ) \\
(1+\beta )\dot{H}=& -4\pi \dot{\phi}^{2}(1+3\kappa H^{2})
\end{align}%
which can be solved as follows 
\begin{align}
\dot{\phi}& =-\frac{(1+\beta )^{3/2}V^{\prime }(\phi )}{2\sqrt{6\pi V(\phi )}%
(1+\beta +8\pi \omega ^{2}V(\phi ))}  \label{a3} \\
\dot{H}& =-\frac{V^{\prime 2}(1+\beta )}{6V(\phi )(1+\beta +8\pi \omega ^{2}
V(\phi ))} \\
H^{2}& =\frac{8\pi }{3(1+\beta )}V(\phi ).
\end{align}%
Using the slow-roll analysis,  the relevant  parameters are found to be 
\begin{align}
\epsilon & =\frac{(1+\beta )^{2}V^{\prime 2}}{16\pi V{(\phi )}^{2}(1+\beta
+8\pi \omega ^{2}V(\phi ))} \\
\eta & =\frac{(1+\beta )^{2}\left( (1+\beta +12\pi  V(\phi\omega ^{2} )){%
V^{\prime }(\phi )}^{2}-V(\phi )(1+\beta +8\pi \omega ^{2} V(\phi ))V^{\prime
\prime }(\phi )\right) }{4\pi V(\phi )^{2}(1+\beta +8\pi \omega ^{2} V(\phi ))^{2}%
}.
\end{align}

In the inflationary model scenarios, such quantities give the scalar field
values $\phi_E$ at the end of the expansion via the constraint $%
\epsilon(\phi_E)=1$. Moreover, the scalar field at the beginning of
inflation can be determined by exploiting the total logarithmic phase. It
turns out that the number of e-folds associated with the inflation duration
will be needed to handle the  cosmological observables. Usually, it reads as 
\begin{equation}
N = \int_{t_I}^{t_E}{\ H dt } = \int_{\phi_I}^{\phi_E}{\ \frac{H }{\dot{\phi}%
} d\phi } =-\int^{\phi_{E}}_{\phi_I} \frac{8\pi}{(1+\beta)^2}\frac{V(\phi)}{%
V^{\prime }(\phi)} (1+\beta +8\pi \omega ^{2} V(\phi)) d\phi
\end{equation}
where one has used the subscript $I$ and $E$ indicating the parameter values
at the onset  and the  offset time of inflation, respectively. The interesting
gravity models should provide consistent predictions which can be  either refuted or
corroborated by the observational data. To give such an evidence, the
inflationary observables should be determined. Following \cite{r29,r30,r32},
the scalar spectral index $n_s$ and the tensor-to-scalar ratio $r$ are  expressed as follows \begin{eqnarray}
n_{s} -1& =& -2\epsilon - \eta \\
r &=& 16 \epsilon.
\end{eqnarray}
In the presence of the kinetic term in the scaled gravity, $n_s$ and $r$ are
modified as follows 
\begin{eqnarray}  \label{a5}
n_{s} &=&1-\frac{(\beta +1)^2 V^{\prime 2}}{8 \pi V(\phi )^2 \left(1+\beta
+8 \pi \omega ^2 V(\phi )\right)} \\
&+&\frac{(\beta +1)^2\left(1+\beta +12 \pi \omega ^2 V(\phi )\right) \left(
V(\phi ) V^{\prime \prime }(\phi ) -V^{\prime 2 }\right)}{4 \pi V(\phi )^2
\left(1+\beta +8 \pi \omega ^2 V(\phi )\right)^2}  \notag \\
r &=& \frac{(\beta +1)^2 V^{\prime 2}}{\pi V(\phi )^2 \left(1+\beta +8 \pi
\omega ^2 V(\phi )\right)}.
\end{eqnarray}
These relations go beyond the known ones. Taking $\beta=0$, we recover the
scalar spectral index $n_s$ and the tensor-to-scalar ratio $r$ of the model
constituting of a scalar field kinetically coupled to a standard gravity
model with a positive coupling constant \cite{r28,r31}. Considering 
$\beta=\omega=0$, we obtain the relations associated with standard inflation.

\section{ Decoupling and coupling  scenarios in scaled gravity inflation }

In this section, we would like to investigate inflation  coupling scenarios in a scaled
gravity by means of   the inflation moduli space. A close examination shows that the
moduli space $\mathcal{M}$ of this theory can be split as follows 
\begin{equation}
\mathcal{M}=\mathcal{M}_{g}\times \mathcal{M}_{m}
\end{equation}%
where $\mathcal{M}_{g}$ is the moduli subspace associated with the gravity
depending on the form of $F(R)$. However, the moduli subspace $\mathcal{M}%
_{m}$ is coordinated by the parameters appearing in the matter sector
described by $\mathcal{L}_{m}$.  Precisely, it depends on $\omega $ and the parameters
of the potential $V(\phi )$. A generic moduli space could generate complex
computations. Here, however, we pay attention to special forms of the
gravity function $F(R)$ and the scalar potential $V(\phi )$. Precisely, we
consider the situation where the gravity function takes the form $%
F(R)=(1+\beta )R$ and the potential is given by 
\begin{equation}
V(\phi )=M^{4}\left[ 1-\cos \left( \frac{\phi }{\mu}\right) ^{\alpha }\right] 
\label{po1}
\end{equation}
where the exponent $\alpha $  is a free parameter.  $M$  and $\mu$ are two free mass scale parameters associated with
the moduli sub-space factor $\mathcal{M}_{m}$ \cite{M4}. These
parameters will be investigated later on. It is worth noting that, for $%
\alpha =1$, the scalar potential reduces to the form
\begin{equation}
V(\phi )=M^{4}\left[ 1-\cos \left( \frac{\phi }{\mu}\right) \right] 
\label{po1}
\end{equation}%
which has been largely studied in connections with particle physics dealing with  axions and DM 
from models going   beyond SM. For generic
values of $\alpha $, we consider two inflation scenarios. In the first one,
the relevant parameters of $\mathcal{M}_{g}$ and $\mathcal{M}_{m}$ are not
bridged and linked. We refer to such a scenario as a  parameter decoupling
 scenario   assured by 
\begin{equation}
\alpha \neq \beta .  \label{alphai}
\end{equation}%
The second scenario corresponds to the case where the relevant parameters of 
$\mathcal{M}_{g}$ and $\mathcal{M}_{m}$ are linked. We refer to such a road
as parameter coupling scenario. A special situation can be occurred by
considering 
\begin{equation}
\alpha =\beta .  \label{alphabeta}
\end{equation}%
The implementation of the gravity parameter in the matter sector can be
considered as a new way to generate an inflation coupling   scenario by means
of the moduli space $\mathcal{M}$.    Precisely, this implementation  could be  interpreted as an alternative way  to generate the coupling between the gravity and the scalar field  via the potential.  Coupling  $ F(R)$ with  such a scalar potential,  the relevant  cosmological quantities get modified   in the same time by  changing the
gravity parameter  $\beta$. In this regard,  the modified gravity controlled by changing $\beta$ in the scalar  potential could provide   models going beyond other investigations where  the gravity  and the scalar  potential are not linked. We anticipate that this   scenario   could
provide  some  models going beyond the previous ones which could be either
rejected or corroborated by experimental findings via   the falsification analysis.

\subsection{Parameter decoupling scenario}

\begin{figure}[!ht]
\centering
\begin{tabbing}
	
		\hspace{5.6cm}\= \hspace{5.6cm}\=\kill
		
		\includegraphics[scale=0.60]{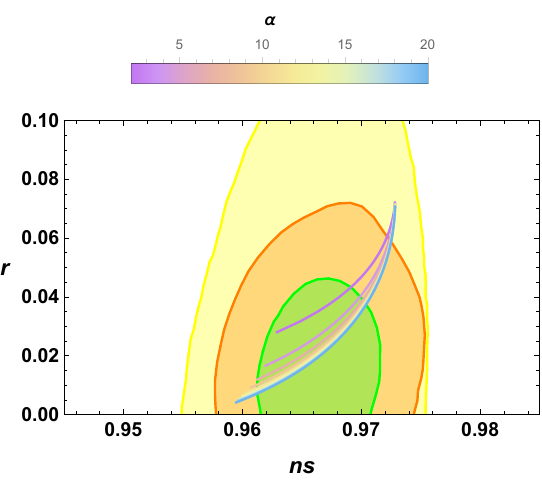} \>
		
		\includegraphics[scale=0.60]{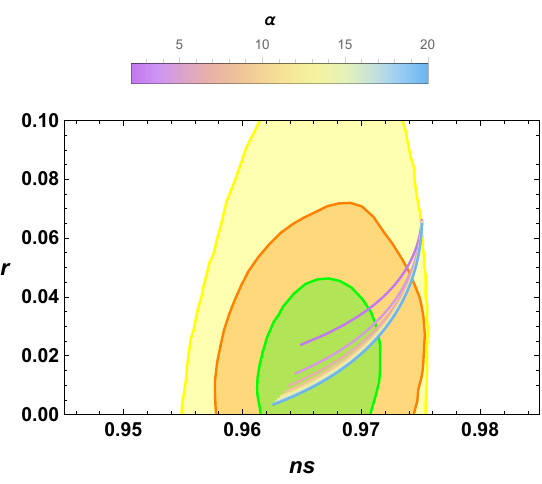} \>
		
		\includegraphics[scale=0.60]{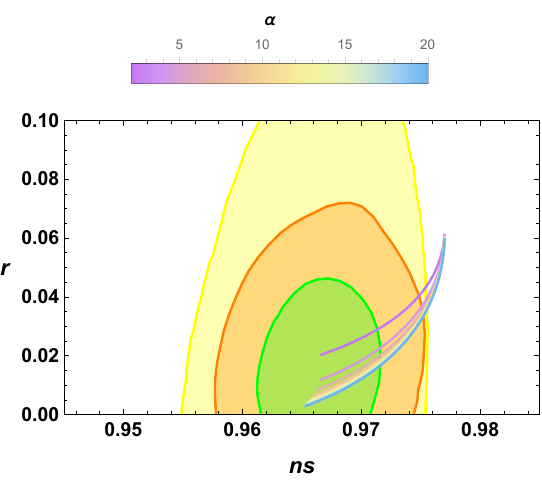} \\
		
	\end{tabbing}
\par
\vspace{-1.5 cm}
\caption{\textit{\protect\footnotesize The behavior of one-dimensional real $%
n_s-r$ curves by taking   $\protect%
\alpha=2.1,\cdots, 20$, the parameter $\protect\omega=1,\cdots, 100$  normalized by the mass scale $M$.  The
left, the center and the right plots correspond to the e-folding number $%
N=55 $, $N=60$ and $N=65$, respectively. The green and the orange contour
constraints represent 68\% and 95\% confidential levels of Planck results
(TT,TE,EE+lowE +lensing +BK15+BAO), respectively. The yellow contour is
associated with the Planck results (TT,TE,EE+lowE+lensing).}}
\label{f3}
\end{figure}

In this subsection, we consider the first inflation scenario. Concretely,
we compute the relevant observables  being the scalar spectral index $n_s$ and
the tensor-to-scalar ratio $r$ as functions of the moduli space coordinates.
To perform such calculations, the number of e-folds $N$ should be
determined. Indeed, it can be expressed as 
\begin{eqnarray}
N=\frac{8 \pi \mu  \, (A_E-A_I)}{\alpha (\beta +1)^2}
\end{eqnarray}
where one has used 
\begin{eqnarray}
A_i&=&-\frac{ \mu (1+\beta +8 \pi M^4 \omega^2) \cos ^{2-\alpha }\left(\frac{%
\phi_i}{\mu  }\right)}{\alpha -2}F_1+\frac{8 \pi \mu  M^4 \omega^2 \cos
^{\alpha +2}\left(\frac{\phi_i}{\mu }\right)}{\alpha +2}F_2  \notag \\
&+&\frac{1}{2} \mu  \left( 1+\beta +16 \pi M^4 \omega^2\right) \log
\left(\sin ^2\left(\frac{\phi_i}{\mu  }\right)\right).
\end{eqnarray}
It  is denoted that $i=E,I$ indicate the onset and  the offset on the  inflationary phase. $F_1$ and 
$F_2$ are the hypergeometric functions reading as 
\begin{eqnarray}
F_1= \, _2F_1\left(1,1-\frac{\alpha }{2};2-\frac{\alpha }{2};\cos ^2\left(%
\frac{\phi_i}{\mu }\right)\right) \\
F_2=\, _2F_1\left(1,\frac{\alpha }{2}+1;\frac{\alpha }{2}+2;\cos ^2\left(%
\frac{\phi_i}{\mu }\right)\right).
\end{eqnarray}
It has been observed that the number of e-folds imposes extra conditions
on $\alpha$. For the present scaled gravity, the scalar spectral index $n_s$
and the tensor-to-scalar ratio $r$ are found to be 
\begin{equation}
\begin{split}  \label{mu1}
n_s&= 1-\frac{\alpha ^2 (\beta +1)^2 \sin ^2\left(\frac{\phi }{\mu  }\right)
\cos ^{2 \alpha -2}\left(\frac{\phi }{\mu }\right)}{8\pi \mu  ^2 \left(1-\cos
^{\alpha }\left(\frac{\phi }{\mu  }\right)\right)^2 \left(1+\beta +8 \pi M^4
\omega^2 \left( 1 -\cos ^{\alpha }\left(\frac{\phi }{\mu  }\right)
\right)\right)} \\
+&\frac{\alpha (\beta +1)^2 \, B \,\cos ^{\alpha -2} \left(\frac{\phi }{\mu  }\right)%
}{8\pi \mu ^2 \left(1-\cos ^{\alpha }\left(\frac{\phi }{\mu  }\right)\right)^2
\left(1+\beta +8 \pi M^4 \omega^2 \left(1-\cos ^{\alpha }\left(\frac{\phi }{%
\mu }\right)\right)\right)^2}  \notag \\
r&=\frac{\alpha ^2 (\beta +1)^2 \sin ^2\left(\frac{\phi }{\mu  }\right) \cos
^{2 \alpha -2}\left(\frac{\phi }{\mu  }\right)}{\pi \mu  ^2 \left(1-\cos
^{\alpha }\left(\frac{\phi }{\mu }\right)\right)^2 \left(1+\beta +8 \pi M^4
\omega^2 \left( 1 -\cos ^{\alpha }\left(\frac{\phi }{\mu  }\right)
\right)\right)}
\end{split}%
\end{equation}
where the quantity $B$ is given by 
\begin{eqnarray}
B&=&\left( 1+\beta +8 \pi M^4 \omega^2 \right) \left(2-\alpha \left(1-\cos
\left(\frac{2 \phi }{\mu  }\right)\right)\right)  \notag \\
&-&2 \cos ^{\alpha }\left(\frac{\phi }{\mu  }\right) \left(1+\beta +16 \pi
M^4 \omega^2-2 \pi \alpha M^4 \omega^2 \left(1-\cos \left(\frac{2 \phi }{\mu  }%
\right)\right)\right)  \notag \\
&+&4 \pi M^4 \omega^2 \left(4+\alpha \left(1-\cos \left(\frac{2 \phi }{\mu }%
\right)\right)\right) \cos ^{2 \alpha }\left(\frac{\phi }{\mu  }\right).
\end{eqnarray}
To validate the obtained results, one should make contact with the
observational findings including  the Planck 2018 and the recently released
BICEP/Keck data\cite{r20,r21,r22}.    Here, we have used a normalized $ \omega$ by multiplying this quantity with $M$ which is fixed to 1.  In Fig(\ref{f3}), we illustrate the $n_s-r
$ curves for different values $N$, $\beta$, $\omega$ and taking  $\mu=M=1$. Instead of giving generic situations, we consider a 
particular one corresponds to $\beta=1$. 
\newline
The associated $n_s-r$ curves are plotted with the presence of the Planck
contour constraints \cite{r20,r21}. In graphic representations, the values
of the number of e-folds have been considered by taking into account of the
experimental constraint  namely $N>60$. In this way, the $n_s-r$ curves have been
analyzed by varying $\alpha$ in the interval $[2.1,20]$. It has been
observed from this figure that the range of $r$ increases by increasing $%
\alpha$. Considering large values of $\alpha$, the range of $n_s$ increases.
Increasing $\omega$, the range of the scalar spectral is increased and the
tensor-to-scalar ratio involves small values. Fixing $\alpha$ and $\omega$, $%
n_s$ increases with $N$. A close inspection reveals that the decoupling
between the scalar potential and the scaled gravity with the kinetic term
could bring interesting numerical results of the spectral index $n_s$ and the
tensor-to-scalar ratio $r$. However, it is not good enough  with respect to the range
associated with the Planck  and the BICEP/Keck data \cite{r20,r21,r22}. The
obtained range of $r$ is $[0.02,0.08]$ which is in a good range but still
not good enough.

\subsection{Parameter coupling scenario}

In this subsection, we follow the second scenario by implementing the
gravity in the matter sector by means of  the scalar  potential. In this way, it takes the following form 
\begin{equation}
V(\phi )=M^{4}\left[ 1-\cos \left( \frac{\phi }{\mu}\right) ^{\beta }\right].
\label{po2}
\end{equation}%
This parameter interplay can be viewed as an alternative way to generate a 
coupling via the moduli space. This provides a bridge between $\mathcal{M}%
_{g} $ and $\mathcal{M}_{m}$ in order to find  results  which could be confronted with
the observational findings. A rapid examination reveals that the scalar
potential form imposes  extra conditions on the gravity parameter $\beta$. It
should be different to $2$ and $-2$. Instead of repeating the computations
and relations, we give only graphic representations.

\begin{figure}[!ht]
\centering
\begin{tabbing}
	
		\hspace{5.6cm}\= \hspace{5.6cm}\=\kill
		
		\includegraphics[scale=0.60]{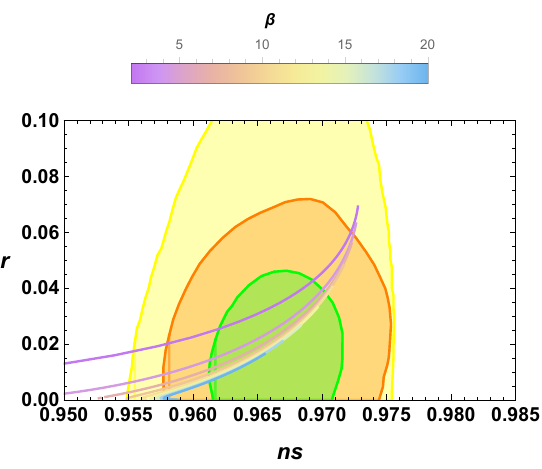} \>
		
		\includegraphics[scale=0.60]{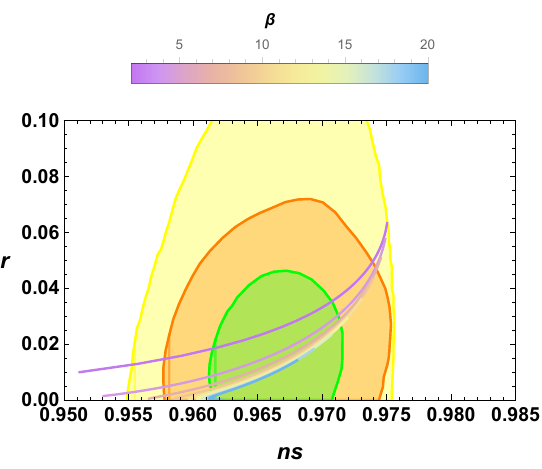} \>
		
		\includegraphics[scale=0.60]{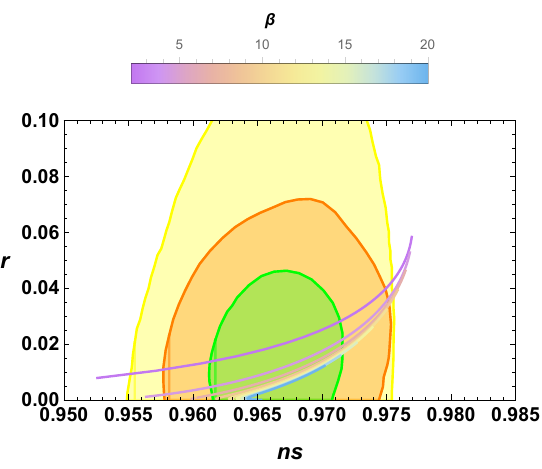} \\
		
	\end{tabbing}
\par
\vspace{-1.5 cm}
\caption{\textit{\protect\footnotesize The behavior of one dimensional
curves $n_s-r$ by varying the gravity parameter $\protect\beta=2.1,\ldots, 20$ and
$\protect\omega=1,\ldots, 100 $  normalized by the mass scale $M$. The plots from left to right
correspond to the number e-folding $N=55$, $N=60$ and $N=65$, respectively.}}
\label{f2}
\end{figure}

As the previous scenario, to check the obtained results, the contact with
observational data including  the Planck 2018 and the recently released
BICEP/Keck data should be provided\cite{r20,r21,r22}.     In Fig(\ref{f2}), we
illustrate the $n_s-r$ curve behaviors by varying $N$, $\beta$, $\omega$,
with    normalized parameters namely $\mu=M=1$. 
\newline
These behaviors are plotted by implementing the Planck contour constraints 
\cite{r20,r21}. They have been examined by varying $\beta$ in the interval $%
[2.1,20]$. It follows from this figure that $r$ decreases by increasing $%
\beta$. Taking large values of $\beta$, the range of $n_s$ decreases.
Increasing $\omega$, the range of the scalar spectral increases and the
tensor-to-scalar ratio takes high values. Fixing $\beta$ and $\omega$, $n_s$
increases by increasing $N$. A close examination shows that the coupling
between the scalar potential and the scaled gravity with a kinetic term
provides interesting numerical values of the spectral index $n_s$ and the
tensor-to-scalar ratio $r$ which are in a  very  good range of the Planck data
which covers both $95\%$ and $68\%$ CL contour regions of the recent
released BICEP/Keck data \cite{r20,r21,r22}. The range of $r$ is
[0.001,0.07] which is in a  very good agreement with the Planck findings being
smaller than $0.1$. It has been anticipated that the coupling between the
scalar  potential  and the scaled gravity with a  kinetic term could be
worked out to bring very good agreements with the experimental data\cite{r30,r31,r32}.

It has been suggested that the scalar power spectrum could be
exploited to discuss the viability of the studied gravity theory \cite{r28}. In
the slow-roll approximations, this quantity takes the form 
\begin{equation}  \label{P1}
P_{\zeta}\approx \frac{H^2}{8\pi^2\epsilon}.
\end{equation}
Taking $\phi=0.1$, $M=\mu=1$, $\omega=0.1$, $\beta=2.1$, this found to be 
\begin{equation}  
	\label{P2}
P_{\zeta}\approx 3.97444. 10^{-5}
\end{equation}
 being an acceptable value.  Other values could be also obtained in certain regions of  the moduli space $\cal M$.

\section{Axionic dark matter}

The idea that the inflaton might be related to axion-like particles belongs
to one of the primary theoretical problems of the minimal inflationary
scenario \cite{38,39,40,41,43,45}. Such a connection is a hint of non ordinary
physics beyond the minimal model of inflation introduced above. Similar to
the axion field suggested to solve the strong CP problem of quantum chromo
dynamics \cite{46,47}, a possible connection is assuming that the inflaton
is a pseudo-Nambu-Goldstone boson (PNGB), of a spontaneously broken
approximate global $U(1)$ symmetry, which arises in the action with a
derivative term $\left( \partial _{\mu }\phi \right) ^{2}$. In fact, it
posses a shift symmetry

\begin{equation}
\phi \rightarrow \phi +c  \label{ax1}
\end{equation}%
where $c$ is a real constant. Such a symmetry saves its role as inflaton
from being invalid  via a coupling to unknown UV physics by severely bordering
the form of its possible interactions with other fields. However, this
continuous symmetry is broken by the ALP potential $V(\phi )$ acquired through
non-perturbative effects   certain  gauge fields $F_{i}$.  They are naturally
coupled to the axion via $\sim g_{i}f_{a}^{-1}\phi F_{i}\overset{\sim }{F_{i}%
}$ with $g_{i}$ is a model-dependent coupling constant, and $f_{a}$ is the axion
decay constant. The latter  determines  the scale at which the axion symmetry is broken.
Supposing the associated global symmetry is spontaneously broken at a scale $%
f_{a}$, with a  soft explicit symmetry breaking at a lower scale $M$, the
axionic inflation is quietly described by these two scales which will be
specified by the exigencies of infallible inflation. Roughly, in connection
with the considered model associated with Eq.(\ref{po2}), the resulting
scaled gravity axion potential is generally of the form

\begin{equation}
V(\phi )=M^{4}\left[ 1-\cos \left( \theta \right) ^{\beta }\right]
\label{ax2}
\end{equation}%
where we have used $\mu=  f_{a}$ and  the canonically normalized field $\phi \equiv f_{a}\theta 
$  where  the dynamical field is constrained by  $\theta \ll 0$ \cite{43,46}. Concretely, the
axion-like field is chosen such that the corresponding potential given by  Eq.(\ref%
{ax2}) is minimized at $\theta =0$. This potential breaks the global shift
symmetry of the axion Eq.(\ref{ax1}) down to a discrete symmetry

\begin{equation}
\phi \rightarrow \phi +2\pi f_{a}.  \label{ax3}
\end{equation}%
By expanding the  scalar potential given by  Eq.(\ref{ax2}) to leading order in $\theta $ like

\begin{equation}
V(\phi )\simeq M^{4}\left[ 1-\left( 1-\frac{\theta ^{2}}{2}\right) ^{\beta }%
\right] \simeq M^{4}\left( \beta \frac{\theta ^{2}}{2}\right) =\frac{1}{2}%
\frac{M^{4}\beta }{f_{a}^{2}}\phi ^{2},  \label{ax4}
\end{equation}%
one can read the axion mass from the only appearing gravity scaled mass term
such as

\begin{equation}
m_{\phi }\simeq \sqrt{\beta }\frac{M^{2}}{f_{a}}.  \label{ax5}
\end{equation}%
It is denoted that the inflation mass scale $M$ and the axion decay constant 
$f_{a}$ are not fixed by theory. This could be scaled by the gravity
parameter $\beta $. Therefore, higher values of the associated axion
symmetry scale $M\ll $ $f_{a}\ll M_{Planck}$ imply lighter and less
interacting axions with the SM. In this case, the corresponding axions would
be relativistic particles. This sets  a  three-dimensional parameter space on
which the axion-like searches depend

\begin{equation}
\mathcal{M}_{\phi }=\{M,f_{a},\beta \}.  \label{ax6}
\end{equation}%
Such weakly interacting light particles could make or contribute as a
sub-component to hot DM in the Universe. Indeed, in such a case, these
particles should have a local mass density of that pretended in our
proximity to account for the dynamics of our galaxy. Concretely, they should
be distributed in a halo endging our galaxy with a characteristic
relativistic velocity close to the  light speed  $v_{a}\sim c$. In spite of
the fact that the axion decays are still constrained by several cosmological
arguments, these particles could be detected either indirectly via their
self-annihilation products likely into photons or neutrinos

\begin{equation}
\phi \phi \rightarrow XX\text{ \ \ \ \ \ \ \ }X=\gamma, \nu 
\label{ax7}
\end{equation}
or directly by considering their interaction via the tiny shocks with the
detector materials. In particular, the corresponding typical kinetic energy
would of the order of

\begin{equation}
K_{\phi }\sim \sqrt{\beta }\frac{M^{2}}{f_{a}}c^{2}\sim eV  \label{ax8}
\end{equation}%
where the axion mass upper bound $m_{\phi }<0.5eV$ is put from the
constrained value of hot DM density by cosmological observations \cite{47}.
In this  way, we can now deal  with the involved inflation scale $M$ of the proposed gravity
model. Concretely, owing to the high and wide energy range of the decay
constant of the axion  $f_{a}\gtrsim 10^{10}GeV$ and according to the
considered range of  the scaled gravity parameter $\beta \leq 20$, we get

\begin{equation}
M\ll f_{a}.  \label{ax9}
\end{equation}%
In this approach, for the  underlying scale $M$
to be high enough to account for inflation, the axionic scale can go up to
the Planck scale $f_{a}\lesssim M_{Planck}$. 

\section{Conclusion}
In this work,    we have  investigated parameter  coupling scenarios in a scaled
gravity via the inflation moduli space.  In particular, we have observed that  this moduli space    contains two factors providing two inflation scenarios. The first one is  called   parameter decoupling scenario  while the second one is parameter  coupling scenario.  Motivated by the modified gravity theories $F(R)\neq R$ and inflationary physics, we   have  proposed and investigated  an inflation model in a scaled
gravity $F(R)=R\,+\beta R$, where $\beta $ is   a  dimensionless scaling
parameter for both scenarios.   For the second one,  this  gravity parameter   has  been implemented in  particular inflation  potential given by  $V(\phi )=M^{4}\left[ 1-\cos \left( \frac{\phi }{\mu}\right)
^{\beta }\right] $  being considered to drive the inflation.  Exploiting  the slow-roll analysis, the gravity scale parameter $\beta $  has  been  approached  with respect to the range of the associated computed cosmological
observables $n_{s}$ and $r$ according to the recent Planck and BICEP/Keck
data. In the second part of this work,  we have investigated an   axionic dark matter in the proposed  gravity  model by considering 
the case where the inflaton  is taken to  be identified with  an axion-like
field $\phi =f_{a}\theta $ with  the decay constant $ f_{a}=\mu$.   Based on  known data,  we have shown that the underlying inflation scale $M$ is
constrained to be much lower than the associated axion scale  $M\ll f_{a}$.

It has been concluded that connecting inflation with axion-like particles could offer an attractive model that can account for most observed cosmological structures, including DM. This investigation road  is now undergoing an expansion phase and the experimental endeavors are rapidly increasing in intensity as well as diversity to trap DM by probing a large fraction of the axion parameter space. Seen that a discovery in the forthcoming years is not precluded, such a finding would be a breakthrough discovery that could reframe the posterior developments  of particle physics,  astrophysics and cosmology, including other high energies theories.
\section*{Declarations}
 The authors declare that they have no known competing interests or personal relationships
that could have appeared to influence the work reported in this paper.
\section*{Ethical Approval} 
It is not applicable in this article.
 \section*{Competing interests}  
  The authors declare that they have no known competing interests.
  \section*{Authors' contributions}  
  The all authors have
worked on the proposed work.
   \section*{Funding}
    No  fundings are associated with this article.
    \section*{Availability of data and materials}
  No data are associated with this article.

\section*{Acknowledgments}

The authors would like to thank   I. Aamer,  N. Askour,   S. Baddis,  H. Belmahi, M. Benali,  H. El Moumni, Y. Hassouni, M. Oualaid, and M.B. Sedra for
collaborations on related subjects.  AB and SEE  would like to thank their  families for support.


\begin{thebibliography}{99}
\bibitem{r0} A. D. Linde, \textit{Generation Of Isothermal Density
Perturbations In The Inflationary Universe}, JETP Lett. \textbf{40}, (1984)
1333 [Pisma Zh. Eksp. Teor. Fiz. 40, 1984) 496 ].

\bibitem{r00} A.A. Starobinsky, \textit{Robustness of the inflationary
perturbation spectrum to trans-Planckian physics},
PismaZh.Eksp.Teor.Fiz.73:415-418,2001,  JETPLett. \textbf{73} (2001)371.

\bibitem{r1} S.~D.~Odintsov and V.~K.~Oikonomou, \textit{Inflationary $%
\alpha $-attractors from $f(R)$ gravity}, Phys. Rev. D \textbf{94}
(2016)124026, \texttt{arXiv:1612.01126}.

\bibitem{r2} I.~Sawicki and W.~Hu, \textit{Stability of Cosmological
Solution in f(R) Models of Gravity}, Phys. Rev. D \textbf{75} (2007) 127502, 
\texttt{arXiv:astro-ph/0702278}.

\bibitem{r3} S.~Carloni, \textit{Covariant gauge invariant theory of Scalar
Perturbations in $f(R)$-gravity: a brief review}, Open Astron. J. \textbf{3}
(2010)76, \texttt{arXiv:1002.3868}.

\bibitem{r4} T.~P.~Sotiriou, \textit{f(R) gravity and scalar-tensor theory},
Class. Quant. Grav. \textbf{23} (2006) 5117, \texttt{arXiv:gr-qc/0604028}.

\bibitem{r5} S.~D.~Odintsov and V.~K.~Oikonomou,\textit{Unification of
Inflation with Dark Energy in $f(R)$ Gravity and Axion Dark Matter}, Phys.
Rev. D \textbf{99} (2019) 104070, \texttt{arXiv:1905.03496}.

\bibitem{r6} A. H. Guth and P.J.~Steinhardt, \textit{The inflationary
universe}, Scientific American \textbf{250} (1984) 129.

\bibitem{r7} A. H. Guth, \textit{The Inflationary Universe: A Possible
Solution to the Horizon and Flatness Problems}, Phys. Rev. D \textbf{23}
(1981) 356.

\bibitem{r8} A.~D.~Linde, \textit{A New Inflationary Universe Scenario: A
Possible Solution of the Horizon, Flatness, Homogeneity, Isotropy and
Primordial Monopole Problems}, Phys. Lett. B \textbf{108} (1982) 393.

\bibitem{r9} J.~A.~Adams, B.~Cresswell and R.~Easther, \textit{Inflationary
perturbations from a potential with a step}, Phys. Rev. D \textbf{64} (2001)
123514, \texttt{arXiv:astro-ph/0102236}.

\bibitem{r11} X.~Chen, R.~Easther and E.~A.~Lim, \textit{Large
Non-Gaussianities in Single Field Inflation}, JCAP \textbf{06} (2007) 023, 
\texttt{arXiv:astro-ph/0611645}.

\bibitem{r12} A.~G. Cadavid and A.~E.~Romano, \textit{Effects of
discontinuities of the derivatives of the inflaton potential}, Eur. Phys. J.
C \textbf{75} (2015) 589, \texttt{arXiv:1404.2985}.

\bibitem{r13} A.~W.~Beckwith, \textit{How a Randall-Sundrum brane-world
effective potential influences inflation physics}, AIP Conf. Proc. \textbf{%
880} (2007) 1180, \texttt{arXiv:physics/0610247}.

\bibitem{r14} H.~V.~Peiris, D.~Baumann, B.~Friedman and A.~Cooray, \textit{%
Phenomenology of D-Brane Inflation with General Speed of Sound}, Phys. Rev.
D \textbf{76} (2007) 103517, \texttt{arXiv:0706.1240}.

\bibitem{r15} M.~Sami, N.~Savchenko and A.~Toporensky, \textit{Aspects of
scalar field dynamics in Gauss-Bonnet brane worlds}, Phys. Rev. D \textbf{70}
(2004) 123528, \texttt{arXiv:hep-th/0408140}.

\bibitem{M1} A.~Belhaj, M.~Benali, Y.~Hassouni, M.~Oualaid and M.~B.~Sedra, 
\textit{On brane cosmological behaviors of Starobinsky inflationary model},
Int. J. Mod. Phys. A \textbf{37} (2022) 2250043.

\bibitem{M2} A.~Belhaj, Y.~Hassouni, M.~Oualaid and M.~B.~Sedra, \textit{On
stringy inflation potentials}, Mod. Phys. Lett. A \textbf{36} (2021) 2150225.

\bibitem{r16} S.~Nojiri, S.~D.~Odintsov and M.~Sami, \textit{Dark energy
cosmology from higher-order, string-inspired gravity and its reconstruction}%
, Phys. Rev. D \textbf{74} (2006) 046004, \texttt{arXiv:hep-th/0605039}.

\bibitem{r17} E.~D.~Stewart, \textit{Inflation, supergravity and superstrings%
}, Phys. Rev. D \textbf{51} (1995) 684, {\tt arXiv:hep-ph/9405389}.

\bibitem{r18} E.~Witten, \textit{Symmetry Breaking Patterns in Superstring
Models}, Nucl. Phys. B\textbf{258} (1985) 75.

\bibitem{r19} T.~J~Li, J.~L.~Lopez and D.~V.~Nanopoulos, \textit{%
Compactifications of M theory and their phenomenological consequences},
Phys. Rev. D \textbf{56} (1997) 2606, \texttt{arXiv:hep-ph/9704247}.

\bibitem{r23} V.~K.~Oikonomou, \textit{Unifying inflation with early and
late dark energy epochs in axion $F(R)$ gravity}, Phys. Rev. D \textbf{103}
(2021) 044036, \texttt{arXiv:2012.00586}.

\bibitem{r24} B.~Li and J.~D.~Barrow, \textit{The Cosmology of f(R) gravity
in metric variational approach}, Phys. Rev. D \textbf{75} (2007) 084010, 
\texttt{arXiv:gr-qc/0701111}.

\bibitem{r25} K.~Bamba, S.~Nojiri, S.~D.~Odintsov and D.~S\'aez-G\'omez, 
\textit{Inflationary universe from perfect fluid and $F(R)$ gravity and its
comparison with observational data}, Phys. Rev. D \textbf{90} (2014) 124061, 
\texttt{arXiv:1410.3993}.

\bibitem{r26} V.~K.~Oikonomou, \textit{Singular Bouncing Cosmology from
Gauss-Bonnet Modified Gravity}, Phys. Rev. D \textbf{92} (2015) 124027, 
\texttt{arXiv:1509.05827}.

\bibitem{r27} V.~K.~Oikonomou, \textit{A refined Einstein\textendash{}Gauss%
\textendash{}Bonnet inflationary theoretical framework}, Class. Quant. Grav. 
\textbf{38} (2021) 195025, \texttt{arXiv:2108.10460}.

\bibitem{M3} A.~Belhaj, H.~Es-Sobbahi, M.~Oualaid and E.~Torrente-Lujan, 
\textit{Reconstructing slow-roll Scalar-Tensor Gauss-Bonnet single field
inflation from running spectral data}, \texttt{arXiv:2108.11881}.


\bibitem{r20} Y.~Akrami \textit{et al.} \textit{Planck 2018 results. X.
Constraints on inflation}, Astron. Astrophys. \textbf{641} (2020) 10, 
\texttt{arXiv:1807.06211}.

\bibitem{r21} N.~Aghanim \textit{et al.}, \textit{Planck 2018 results. VI.
Cosmological parameters}, Astron. Astrophys. \textbf{641} (2020) 6, \texttt{%
arXiv:1807.06209}.

\bibitem{r22} P.~A.~R.~Ade \textit{et al.}, \textit{Improved Constraints on
Primordial Gravitational Waves using Planck, WMAP, and BICEP/Keck
Observations through the 2018 Observing Season}, Phys. Rev. Lett. \textbf{127%
} (2021) 151301, \texttt{arXiv:2110.00483}.

\bibitem{T1} H. Tiberiu, S. N. L. Francisco,  N. Shin\'ichi and S.D. Odintsov, {\it $f(R,T)$ gravity} Phys. Rev. D  \textbf{84} (2011) 024020,  {\tt arXiv:1104.2669}.
\bibitem{T2} P. H. R. S. Moraes, J.  D.  V. Arba\~nil, and M. Malheiro, {\it Stellar equilibrium configurations of compact stars in $f(R,T)$ gravity},  JCAP  \textbf{06} (2016) 005,   {\tt arXiv:1511.06282}.

\bibitem{T3}  F. G. Alvarenga,  A.  dela Cruz-Dombriz, M. J. S.  Houndjo,  M. E. Rodrigues,  and D. S\'aez-G\'omez, {\it Dynamics of scalar perturbations in $f(R,T)$ gravity}, Phys. Rev. D 10 \textbf{87}  (2013) 103526,  {\tt arXiv:1302.1866}.

\bibitem{T4} H. Velten and  T. R.  P. Caram\^es, {\it Cosmological inviability of $f(R,T)$ gravity}, Phys. Rev. D 95 \textbf{12} (2017) 123536, 
{\tt arXiv:1702.07710}.
\bibitem{T5}
S. Taghavi,  K. Saaidi  and  Z. Ossoulian,
{\it Holographic inflation in $f(R,T)$ gravity},
{\tt arXiv:2301.02631}.

\bibitem{T6} S. Bhattacharjee,  J. R. L. Santos, P. H. R. S. Moraes   and P. K. Sahoo, {\it Inflation in $f(R, T)$ gravity}, Eur. Phys. J. Plus 135 \textbf{7}  (2020) 576,  {\tt arXiv:2006.04336.}




\bibitem{d1} S. Nojiri, S. D. Odintsov, and M. Sasaki, \textit{Gauss-Bonnet
dark energy}, Phys. Rev. D {\bf 71} (2005) 123509, {\tt arXiv:hep-th/0504052}.

\bibitem{d2} S. Nojiri, S. D. Odintsov, and M. Sami, Phys. Rev. D {\bf 74} (2006)
046004, \texttt{arXiv:hep-th/0605039}.
\bibitem{sm1} C. Germani, and K. Kehagias, \textit{New Model of Inflation
with Non-minimal Derivative Coupling of Standard Model Higgs Boson to Gravity%
}, Phys. Rev. Lett. \textbf{105} (2010) 011302, {\tt arXiv:1003.2635}.

\bibitem{sm2} S. Tsujikawa, \textit{Observational tests of inflation with a
field derivative coupling to gravity}, Phys. Rev. D \textbf{85} (2012)
083518, \texttt{arXiv:1201.5926}.

\bibitem{dm1} J.S.  Matthew,  D.J.E. Marsh,  C.  Pongkitivanichkul,  L.C. Price and B.S. Acharya, \textit{Spectrum of the axion dark sector%
},  Phys. Rev. D 96 \textbf{8} (2017)10 \texttt{arXiv:1706.03236}. 

\bibitem{dm2}D.  J. H. Chung, E.  W. Kolb, and A. 
Riotto,  {\it Superheavy dark matter}, Phys. Rev. D {\bf  59}
(1998)023501,  {\tt arXiv:hep-ph/9802238}.
\bibitem{dm3} V.  Kuzmin and I. Tkachev,  \textit{Matter creation via
vacuum fluctuations in the early universe and observed
ultrahigh-energy cosmic ray events}, Phys. Rev. D  {\bf 59},
(1999) 123006,  {\tt arXiv:hep-ph/9809547}.
\bibitem{dm4} A. Bhoonah, J.  Bramante, S.  Nerval, and
N.  Song,   \textit{Gravitational Waves From Dark Sectors,
Oscillating In
atons, and Mass Boosted Dark Matter}, 
JCAP 04(2021) 043 ,  {\tt arXiv:2008.12306}.

\bibitem{dm5}  C. A. Baker, D. D. Doyle, P. Geltenbort, K. Green,
M. G. D. van der Grinten, P. G. Harris, P. Iaydjiev, S. N.
Ivanov, D. J. R. May, J. M. Pendlebury, J. D. Richardson,
D. Shiers, and K. F. Smith, \textit{Improved Experimental
Limit on the Electric Dipole Moment of the Neutron}.
Phys. Rev. Lett.  {\bf 97} (2006)131801.
\bibitem{dm6}  R. D. Peccei and H. R. Quinn, {\it CP Conservation in the
Presence of Pseudoparticles}. Phys. Rev. Lett. {\bf  38}(1997)1440.
\bibitem{dm7} 
A. Arvanitaki, S. Dimopoulos, S. 
Dubovsky, N.  Kaloper, and J.  March-Russell.
{\it String axiverse}. Phys. Rev. D {\bf 81} (2010) 123530.
\bibitem{dm8} 
M. Green, J.H. Schwarz and E. Witten, Superstring Theory, (2 volumes), Cambridge
University Press (1986).
\bibitem{dm9} 
S.  Alexander and N.  Yunes, {\it  Chern–Simons
modified general relativity},  Physics Reports, {\bf 480}(1-2) (2009)
55.
\bibitem{dm10} 
S. Jung, T. Kim, J. Soda, and Y. 
Urakawa, { \it Constraining the gravitational coupling of axion
dark matter at LIGO} Phys. Rev. D {\bf 102} (2020)055013.

\bibitem{r28} N.~A.~Avdeev and A.~V.~Toporensky, \textit{Ruling out an
inflation driven by a power law potential: kinetic coupling does not help},
Gravitation and Cosmology ~\textbf{28}(2022) 416, \texttt{%
arXiv:2203.14599}.

\bibitem{r29} J.~ Matsumoto and S.~V.~Sushkov, \textit{Cosmology with
nonminimal kinetic coupling and a Higgs-like potential}, \texttt{%
arXiv:1510.03264}.

\bibitem{r30} S.~Tsujikawa, \textit{Observational tests of inflation with a
field derivative coupling to gravity}, Phys. Rev. D \textbf{85} (2012)
083518.

\bibitem{r31} N.~Avdeev, A.~Toporensky, \textit{On viability of inflation in
non-minimal kinetic coupling theory}, Gravitation and Cosmology ~ 
\textbf{27} (2021)269, \texttt{arXiv:2103.00556}.

\bibitem{r32} L.~N.~Granda, D.~F.~Jimenez, \textit{Slow-roll inflation with
exponential potential in scalar-tensor models}, Eur. Phys. J. C \textbf{79}
(2019) 772.

\bibitem{ref1}   V. K. Oikonomou, K-R Revis, I.  C. Papadimitriou, M-M Pegioudi,   \textit{Swampland Criteria and Constraints on Inflation in a f(R,T) Gravity Theory},  Inter.  J. of Mod.  Phys.  \textbf{D32} (2023) 2350034,  \texttt{arXiv:2303.14724}.
\bibitem{ref2} A.  Gitsis, K-R Revis, S. A. Venikoudis, F. P. Fronimos,  \textit{Swampland criteria for rescaled Einstein-Hilbert gravity with string corrections},  \texttt{arXiv:2301.08126}.
\bibitem{ref3} V. K. Oikonomou, I. Giannakoudi, A.  Gitsis, K-R. Revis,  \textit{Rescaled Einstein-Hilbert Gravity: Inflation and the Swampland Criteria},   Inter.  J. of Mod.  Phys.  \textbf{D31}  (2022) 2250001, \texttt{ arXiv:2105.11935}.


\bibitem{M4} J. Martin, C. Ringeval and V. Vennin, \textit{Encyclop\ae {}dia
Inflationaris}, Phys. Dark Univ. \textbf{6} (2014) 235, \texttt{%
arXiv:1303.3787}.

\bibitem{38} M. Kowalski et al, {\it Improved Cosmological Constraints from New, Old and Combined Supernova Datasets}, Astrophys. J. \textbf{686} (2008) 749.

\bibitem{39} A G. Riess et al.,  {\it A Redetermination of the Hubble Constant with the Hubble Space Telescope from a Differential Distance Ladder},  Astrophys. J. \textbf{699} ( 2009) 539.

\bibitem{40} R. Amanullah et al., {\it Spectra and Light Curves of Six Type Ia Supernovae at $ 0.511< z < 1.12$ and the Union 2 Compilation},  Astrophys. J. \textbf{716} (2010) 712.

\bibitem{41} J. Martin,  {\it Everything You Always Wanted To Know About The Cosmological Constant Problem (But Were Afraid To Ask)}, Comptes Rendus Physique \textbf{13} (2012) 566.



\bibitem{43} S. Weinberg,  {\it A New Light Boson?}, Phys. Rev. Lett. {\bf 40} (1978) 223.



\bibitem{45} R. D.  Peccei and H.R.  Quinn, {\it  Constraints Imposed by CP
Conservation in the Presence of Instantons}, Phys. Rev. D{\bf16} (1977) 1791.

\bibitem{46} F. Wilczek, {\it  Problem of Strong P and T Invariance in the
Presence of Instantons}, Phys. Rev. Lett. {\bf 40} (1978) 279.

\bibitem{47} E. Di Valentino, E. Giusarma, M. Lattanzi, O. Mena, A.
Melchiorri and J. Silk,  {\it Cosmological axion and neutrino mass constraints
from Planck 2015 temperature and polarization 1489 data}, Phys. Lett.  B  {\bf 752}
 (2016)182.


\end{thebibliography}
\end{document}